\documentclass[a4paper,11pt]{article}
\usepackage{jheppub} 
\usepackage{lineno}
\usepackage[utf8]{inputenc}

\usepackage{xcolor}
\usepackage{subfigure}
\usepackage{booktabs}
\usepackage{graphicx}
\usepackage[percent]{overpic}
\usepackage{tikz}
\renewcommand{\thesection}{\arabic{section}}

\numberwithin{equation}{section}

\title{\boldmath Learning geometries beyond asymptotic AdS}

\author[a]{Cheng Ran,}
\author[a,b]{Shao-Feng Wu,}
\author[c]{and Zhuo-Yu Xian}

\affiliation[a]{Department of Physics, Shanghai University, Shanghai, 200444, China}
\affiliation[b]{Center for Gravitation and Cosmology, Yangzhou University, Yangzhou 225009, China}
\affiliation[c]{Department of Physics, Freie Universit\"at Berlin, Arnimallee 14, DE-14195 Berlin, Germany}
\emailAdd{r\_cheng@shu.edu.cn}
\emailAdd{sfwu@shu.edu.cn}
\emailAdd{zhuo-yu.xian@fu-berlin.de}

\abstract{
We present a data-driven method for holographic bulk reconstruction that works even when the spacetime is not asymptotically AdS. Given the data of boundary Green functions within a finite frequency window, we iteratively adjust a bulk metric with a finite radial cutoff until its holographic Green functions reproduce the boundary data. Based on the holographic Wilsonian renormalization group for the Klein-Gordon equation in an undetermined curve space, we construct a radial flow equation and transform it into a Neural ODE, which is an infinite-depth neural network for modeling continuous dynamics. Assuming the double-trace coupling $h$ in the Wilsonian action is real, we demonstrate that the Neural ODE can effectively learn the metrics with AdS, Lifshitz, and hyperscaling violated asymptotics. In particular, we apply the algorithm to the Sachdev-Ye-Kitaev (SYK) model which slightly deviates from the conformal limit. In the hyperparameter space spanned by the rescaled temperature $\bar{T}$ and the radial cutoff $\epsilon$, we identify a critical curve along which the learned metric is close to AdS$_2$ black hole with finite cutoff. We derive an approximate analytical expression for this curve, from which an effective bulk dual of the SYK coupling $v$ is established. Our work provides a promising way for using machine learning to depict the novel bulk geometry dual to the non-conformal boundary systems in the real world.}

\begin{document} 
\maketitle
\flushbottom

\section{Introduction}

The AdS/CFT correspondence posits an equivalence between a $(d+2)$-dimensional super gravity in Anti-de Sitter (AdS) space and a $(d+1)$-dimensional super conformal field theory (CFT) \cite{Maldacena:1997re, Witten:1998qj, Gubser:1998bc, Aharony:1999ti}. 
Numerous deformations of this duality have been explored. Certain deformations breaking or preserving the super symmetry drive the system away from the original CFT, initiating a renormalization group (RG) flow \cite{Leigh:1995ep, Freedman:1999gp, Khavaev:1998fb}. The extra radial direction on the gravity side is interpreted as the energy scale in the RG flow of the dual field theory \cite{deBoer:1999tgo,deBoer:2000cz, Bianchi:2001kw, Skenderis:2002wp}, where the small (long)  distance in the field theory corresponds to the boundary (center) of the spacetime, the namely the IR/UV connection \cite{Susskind:1998dq}. 
Such an interpretation allows us to discuss the duality between the gravity theory within a finite IR cutoff surface and the boundary theory with a finite UV cutoff. The Wilson RG connecting the boundary theories defined at different UV cutoffs could be interpreted as integrating out the geometry between the cutoff surfaces on the gravity side \cite{Heemskerk:2010hk, Faulkner:2010jy, Bredberg:2010ky, Nickel:2010pr, Radicevic:2011py, Sin:2011yh, Grozdanov:2011aa, Lizana:2015hqb, Sathiapalan:2017frk, Kim:2021qbi, Mandal:2016rmt, Laia:2011wf, Oh:2012bx}.

Based on the property of strongly/weakly coupled duality, the holographic correspondence has provided rich phenomenological models for strongly coupled quantum systems \cite{Hartnoll:2016apf, Liu:2020rrn}. However, applying AdS/CFT to condensed matter theories (AdS/CMT) faces a fundamental challenge: real condensed matter systems are often ``neither relativistic nor conformal" \cite{Anderson2013}, as Anderson pointed out. Similarly, concerns have been raised regarding AdS/QCD duality, with McLerran noting that ``$\mathcal N = 4$ supersymmetric Yang–Mills is not QCD... It has no mass scale and is conformally invariant" \cite{McLerran:2007hz}.

These practical concerns motivate exploring holographic models beyond asymptotic AdS space, which corresponds to the field theory beyond the conformal UV fixed point. Instead, it could exhibit Lifshitz scaling \cite{Kachru:2008yh}, hyperscaling-violating \cite{Dong:2012se}, or even more exotic behaviors without manifest Lorentz symmetry \cite{Taylor:2008tg}.

According to the holographic principle, complete boundary data of a large-$N$ quantum field theory uniquely determine a dual bulk description, allowing one, in principle, to reconstruct both the bulk action and the spacetime geometry. In bottom-up holographic model building, one usually chooses an appropriate bulk action and boundary conditions, then solves for the saddle-point configurations of the metric and matter fields \cite{Zaanen:2015oix, Ammon:2015wua, Hartnoll:2016apf}. 
A more direct approach is to reconstruct the bulk metric from the boundary data, including
(i) boundary $n$-point functions and related features \cite{deHaro:2000vlm,Hubeny:2006yu,Hammersley:2006cp,Bilson:2008ab,Qi:2013caa,Engelhardt:2016wgb,Engelhardt:2016crc, Fan:2023bsz, Hashimoto:2022aso, Caron-Huot:2022lff, Lu:2025pal, Lu:2025jgk}; 
(ii) subregion entropies \cite{Bilson:2010ff, Nozaki:2012zj, Czech:2015qta, Hammersley:2007ab, You:2017guh, Hubeny:2012ry, Roy:2018ehv, Myers:2014jia, Balasubramanian:2013lsa,Czech:2014ppa, Jokela:2023rba, Jokela:2020auu}; 
(iii) complexity \cite{Xu:2023eof, Hashimoto:2021umd}; (iv) Wilson loop \cite{Hashimoto:2020mrx}; (v) modular Hamiltonians of boundary subregions \cite{Kabat:2018smf}. Beyond the reconstruction of the metric itself, one can also reconstruct local bulk operators explicitly in terms of boundary operators 
\cite{Hamilton:2005ju,Dong:2016eik}.

The spacetime reconstructed from the boundary data of the field theory without a conformal UV fixed point should be naturally go beyond asymptotic AdS. For instance, recent quantum circuit constructions based on generalized free fields suggest that the Sachdev-Ye-Kitaev (SYK) model retains a geometric bulk description even beyond the conformal limit, although the curvature of the emergent spacetime may diverge near the boundary \cite{Nebabu:2023iox}.

However, practical limitations often restrict the available data. The constraints of measurement techniques in experiments, limited storage resources and processing capabilities mean that the spectrum data collected from a real boundary system cannot cover all frequencies. As a result, the artificial UV cutoff would be introduced in the spectrum data, which prevents the exact reconstruction of the asymptotic structure of the bulk spacetime and may introduce systematic errors. 

To address these challenges, we introduce a novel machine learning algorithm that circumvents the reliance on AdS asymptotics. Deep learning (DL) offers a promising avenue for reconstructing bulk metrics. Leveraging the powerful feature extraction capabilities and deployment flexibility of neural networks, this data-driven approach, known as the AdS/DL correspondence \cite{Hashimoto:2018ftp}, has been applied to holographic modeling of QCD and CMT \cite{Hashimoto:2018bnb, Akutagawa:2020yeo, Hashimoto:2021ihd, Yan:2020wcd, Gu:2024lrz, Li:2022zjc, Hashimoto:2020jug, Hashimoto:2022eij, Ahn:2024gjf, Luo:2024iwf, Chen:2024ckb, Chen:2024mmd, Cai:2024eqa, Mansouri:2024uwc, Ahn:2024jkk,  Chen:2024epd, Ahn:2025tjp, Hashimoto:2024yev}. 

In this paper, our approach utilizes the holographic Wilsonian renormalization group (HWRG) proposed in \cite{Faulkner:2010jy} as the theoretical framework. HWRG provides a pathway to a semi-holographic effective action \cite{Jensen:2011af} where the IR is described by dynamical bulk fields, and the UV arises from integrating out the bulk geometry, equivalent to integrating out high-energy modes in the field theory. This process generates a Wilsonian effective action that dictates the boundary conditions for the bulk fields. As pointed out in \cite{Faulkner:2010jy}, HWRG can be applied to various asymptotic geometries, although it does not inherently produce the geometry itself due to the imposed probe limit. In contrast, AdS/DL is constrained by its asymptotics but excels at extracting metrics from field theory data. The core idea of this paper is that their combination provides a powerful and complementary method for constructing bulk geometries beyond asymptotic AdS.

The rest of this paper is organized as follows: In section \ref{HWRG}, we present the ansatz for the action according to HWRG and derive the Green functions of scalar operators in alternative quantization. Section \ref{RG} constructs a general radial flow equation for the bulk response function, which is then translated into the Neural ODE in section \ref{ML}. Additionally, we introduce the loss function and inductive bias for training the Neural ODE. Section \ref{epl} applies the algorithm to the typical models which have different asymptotic geometries and the SYK
model which can deviate from the conformal limit. We provide conclusions and discussions in section \ref{Disc}. There are also three appendices. App.~\ref{stdc} introduces the Green function in standard quantization. App.~\ref{KL} explains the field-theory motivation of our loss function. The algorithm details and training report are given in App.~\ref{AD}.

\section{Holographic Wilsonian renormalization group }\label{HWRG}

We will first review the HWRG following ref.~\cite{Faulkner:2010jy} in order to define a holographic theory with a finite UV cutoff. 

Consider that a quantum system described by quantum field theory with temporal and spatial translational invariance and rank-$N$ gauge fields collectively denoted by $\Phi$. The partition function can be defined by a path integral below some UV cut-off $\Lambda$
 \begin{equation}
Z=\int_{\Lambda }D\Phi \exp \left( iI_{0}[\Phi ]+iI_{\mathrm{UV}}[\Phi
]\right) ,  \label{ZQFT}
\end{equation}
where $I_{0}$ is the original action and $I_{\mathrm{UV}}$ arises from integrating out the degrees of freedom above $\Lambda $. 
Given the one and two-point functions of a single-trace scalar operator $O$ near equilibrium, one can approximate the partition function with its source $J$ at quadratic order as
\begin{equation}\label{pfft}
Z\left[ J\right] \approx Z_{0}\exp \left\{ i\int^\Lambda \frac{d^{d+1}k}{\left( 2\pi \right) ^{d+1}}\left[ J(k)\left\langle
O(-k)\right\rangle +\frac{1}{2}J(k)G(k)J(-k)\right] \right\}.
\end{equation}
For simplicity, we consider the case of a vanishing one-point function, $\langle O\rangle = 0$, which can always be achieved by redefining the operator as $O \to O - \langle O\rangle$.

Assuming the above quantum field theory is strongly coupled in the large-$N$ limit, we expect it to enjoy a holographic dual described by semiclassical gravity, with bulk fields corresponding to single-trace operators. In this paper, we consider only boundary systems that are static, spatially homogeneous, and isotropic when $J=0$. To describe the quadratic generating function of  $J$ in \eqref{pfft} from the bulk point of view, we consider free fields living on the $(d+2)$-dimensional bulk geometry
\begin{equation}\label{ds2}
ds^{2}=-g_{tt}\left( r\right) dt^{2}+g_{rr}\left( r\right)
dr^{2}+g_{xx}\left( r\right) d\vec{x}^{2},
\end{equation}
which is also static, spatially homogeneous and isotropic. We consider that the metric ansatz admits a horizon at $r=r_h$ where $g_{tt}(r_h)=1/g_{rr}(r_h)=0$ and $g_{xx}(r_h)$ is finite.
For simplicity, we will focus on one scalar field $\phi$ in the bulk, which is dual to the scalar operator $O$ on the boundary. The action of the free field $\phi$ can be written as
\begin{equation}\label{ttS}
S=S_{0}[\phi ,r<r_{\epsilon }]+S_{\mathrm{B}}[\phi ,r_{\epsilon }],
\end{equation}
where $S_{0}$ is the bulk action
\begin{equation}\label{S0}
S_{0}=\int_{r<r_{\epsilon }}drd^{d+1}x\sqrt{-g}\left[ -\frac{1}{2}\left(
\partial \phi \right) ^{2}-\frac12 m^2\phi^2 \right]
\end{equation}
and $S_{\mathrm{B}}$ is a boundary action defined at the cutoff surface $%
r=r_{\epsilon }$. In HWRG, $S_{\mathrm{B}}$ is interpreted as dual to $I_{\mathrm{UV}}$ in (\ref{ZQFT}). In alternative quantization, the boundary value of $\phi $ is identified as the expectation value of the scalar operator, and $S_{\mathrm{B}}$ can be directly translated into $I_{\mathrm{UV}}$ up to some renormalization. 
To derive the one and two points functions from the bulk, we only need to expand $S_{\mathrm{B}}$ in quadratic order of $\phi$. In momentum space,
\begin{equation}
S_{\mathrm{B}}=\Lambda \left( r_{\epsilon }\right) +\int_{r=r_{\epsilon }}%
\frac{d^{d+1}k}{\left( 2\pi \right) ^{d+1}}\sqrt{-\gamma }\left[ J(k)\zeta
\phi (-k,r_{\epsilon})+\frac{1}{2}h(k)\phi (k,r_{\epsilon})\phi (-k,r_{\epsilon})\right] ,  \label{SBJh}
\end{equation}
where $\zeta $ is the renormalization factor,  $J$ is the source, and $h$ is the coupling of double-trace deformation.

Because the bulk theory is quadratic in $\phi$, we can compute the two-point function by working at the saddle point $\phi_c$ of the total bulk action.
\begin{equation}
0=\left. \frac{\delta \left( S_{0}+S_{\mathrm{B}}\right) }{\delta \phi }%
\right\vert _{\phi =\phi _{c}}
\end{equation}
is determined by the KG equation and the Neumann boundary condition
\begin{eqnarray}
0 &=&\nabla ^{2}\phi -V^{\prime }(\phi )=\frac{1}{\sqrt{-g}}\partial _{r}(%
\sqrt{-g}g^{rr}\partial _{r}\phi )-\left( k^{2}+m^{2}\right) \phi ,\label{KG} \\
0 &=&\left. \sqrt{-g}g^{rr}\partial _{r}\phi +\sqrt{-\gamma }\left( J\zeta
+h\phi \right) \right\vert _{r=r_{\epsilon }},  \label{UVBC}
\end{eqnarray}
with $k^{2}=-g^{tt}\omega ^{2}+g^{xx}\vec{k}^{2}$. For later convenience, the KG equation can be written in the Hamiltonian form
\begin{equation}
\partial _{r}\phi =-\frac{1}{\sqrt{-g}g^{rr}}\Pi ,\quad\partial _{r}\Pi =-\sqrt{%
-g}\left( k^{2}+m^{2}\right) \phi .
\end{equation}
Then the on-shell action becomes
\begin{eqnarray}
S_{c} &\approx &S_{0}[\phi _{c};r<r_{\epsilon }]+S_{\mathrm{B}}[\phi
_{c};J,r_{\epsilon }]  \notag \\
&=&\int \frac{d^{d+1}k}{\left( 2\pi \right) ^{d+1}}\left. \left[ -\frac{1}{2}%
\phi _{c}\Pi _{c}+\sqrt{-\gamma }\left( J\zeta \phi _{c}+\frac{1}{2}\phi
_{c}h\phi _{c}\right) \right] \right\vert _{r=r_{\epsilon }}+\Lambda \left(
r_{\epsilon }\right)  \notag \\
&=&\int \frac{d^{d+1}k}{\left( 2\pi \right) ^{d+1}}\sqrt{-\gamma }\frac{1}{2}%
\left. J\zeta \phi _{c}\right\vert _{r=r_{\epsilon }}+\Lambda \left(
r_{\epsilon }\right) ,  \label{Sc}
\end{eqnarray}
where we have used (\ref{UVBC}) in the last line.

By comparing the generating function \eqref{pfft} and the on-shell action \eqref{Sc}, we can read off the Green function in the frequency-momentum domain
\begin{equation}
G_{a}=\left. J^{-1}\sqrt{-\gamma }\zeta \phi _{c}\right\vert _{r=r_{\epsilon
}}.  \label{Galt}
\end{equation}

The above formulation works in the alternative quantization \cite{Klebanov:1999tb}. In standard quantization, the boundary value of the scalar field is interpreted as the source, and we can obtain the Green function (detailed in App.~\ref{stdc})
\begin{equation}
G_{s}=\left. \frac{1}{\gamma\zeta ^{2}}\left( \frac{\Pi _{c}}{\phi _{c}}-\sqrt{-\gamma }%
h\right) \right\vert _{r=r_{\epsilon }}.  \label{Gsta}
\end{equation}
Note that (\ref{Galt}) and (\ref{Gsta}) depend on general ($\omega ,%
\vec{k}$). However, we will henceforth focus on $\vec{k}=0$, as all the data considered below have $\vec{k}=0$.

\section{Radial flow equation}\label{RG}

Now we will discuss the way to find the on-shell solution. By defining a response function
\begin{equation}
\chi \equiv \frac{\Pi _{c}}{i\omega \phi _{c}},  \label{chi}
\end{equation}
we can recast the KG equation \eqref{KG} as a first-order radial flow equation: 
\begin{equation}
\partial _{r}\chi -i\omega \sqrt{\frac{g_{rr}}{g_{tt}g_{xx}^{d}}}\left( \chi
^{2}-g_{xx}^{d}\right) +\frac{1}{i\omega }\sqrt{g_{tt}g_{rr}g_{xx}^{d}} m^{2}=0.  \label{floweq1}
\end{equation}
At the horizon $r=r_{h}$, the in-falling boundary condition implies \cite{Son:2002sd}
\begin{equation}
\chi \left( r_{h}\right) =\left[ g_{xx}(r_{h})\right] ^{\frac{d}{2}}.
\end{equation}
Integrating the flow equation from the horizon to the UV cutoff gives $\chi
\left( r_{\epsilon }\right) $, which can be related to the retarded Green functions as follows.

Using (\ref{UVBC}) and (\ref{chi}), we can rewrite (\ref{Galt}) as
\begin{equation}
G_{a}=\left. \frac{\gamma \zeta ^{2}}{-i\omega \chi +\sqrt{-\gamma }h} \right\vert _{r=r_{\epsilon }}.  \label{Ga}
\end{equation}
Inserting (\ref{chi}) into (\ref{Gsta}), we also read
\begin{equation}
G_{s}=\left.\frac{i\omega \chi -\sqrt{-\gamma }h}{\gamma \zeta ^{2}}
\right\vert _{r=r_{\epsilon }}\text{.}  \label{Gs}
\end{equation}

Considering a rescaling 
\begin{align}\label{rscaling}
    r\rightarrow r\lambda, \quad \omega \rightarrow \omega \lambda^{z},\quad 
m^{2}\rightarrow
m^{2}\lambda ^{\frac{2\theta }{d}}
\end{align}
with two arbitrary exponents $z$ and $\theta$, 
we notice that the flow equation (\ref{floweq1}) is invariant if the fields follow
\begin{equation}\label{scaling}
g_{tt}\rightarrow g_{tt}\lambda ^{-\frac{2\theta }{d}
+2z},\quad 
g_{rr}\rightarrow g_{rr}\lambda ^{-\frac{2\theta }{d}-2},\quad
g_{xx}\rightarrow g_{xx}\lambda ^{-\frac{2\theta }{d}+2}, \quad  \chi \to \chi \lambda^{-\theta+d}.
\end{equation}
The two exponents will later be identified as the dynamical critical exponent and the hyperscaling-violation exponent, once the metric is specified. To incorporate the case 
$d=0$, we first take the limit $\theta/d\to0$ and then send $d\to0$ in the scaling transformation. We exploit the underlying symmetry to simplify the metric ansatz, making it suitable for numerical analysis.

When $d\geq1$, since the radial coordinate 
$r$ can be freely redefined, only two independent radial functions are required in the ansatz \eqref{ds2}. Without loss of generality, we fix the radial gauge such that
$g_{xx}(r)=r^{-\frac{2\theta}{d}+2}$. Incorporating the scaling symmetry, the metric ansatz then takes the form
\begin{equation}
ds^{2}=r^{-\frac{2\theta }{d}}\left[ -r^{2z}f_{1}(r)dt^{2}+\frac{1}{r^{2}f_{2}(r)}dr^{2}+r^{2}d\vec{x}^{2}\right],  \label{f12}
\end{equation}
where $f_{1}$ and $f_{2}$ are invariant under the scaling transformation \eqref{scaling}. 
To further simplify the subsequent machine learning analysis in parameter space, we restrict to the case
\begin{equation}
f_1 = f_2 = f, \label{f12f}
\end{equation}
and all data used in this paper are chosen to be consistent with this condition.

When $d=0$, the $d\vec{x}^2$ component is absent, so we may adopt a simplified ansatz for numerics,
\begin{align}\label{f2D}
    ds^2=-r^2f(r)dt^2+\frac1{r^2f(r)}dr^2
\end{align}
by imposing the radial gauge. This ansatz is related to \eqref{f12}\eqref{f12f} by dropping the $d\vec{x}^2$ term and setting $\theta/d=0$ and $z=1$.

By taking the horizon coordinate $r_{h}$ as the reference scale, we define the invariant quantities under the scaling transformation \eqref{rscaling}\eqref{scaling}
\begin{equation}
u\equiv \frac{r_{h}}{r},\quad
\bar{\omega}\equiv \frac{\omega }{r_{h}^{z}},\quad
\bar{\chi}\equiv \frac{\chi }{r_{h}^{d-\theta }},\quad 
\bar{m}\equiv \frac{m}{
r_{h}^{\frac{\theta }{d}}},  \label{dimless}
\end{equation}
As a result, the flow equation can be reduced to
\begin{equation}
\partial _{u}\bar{\chi}+u^{z-1}\frac{i\bar{\omega}}{f(u)}\left( u^{d-\theta }\bar{\chi}^{2}-\frac{1}{u^{d-\theta }}\right) -\frac{1}{i\bar{\omega}}\frac{\bar{m}^{2}}{u^{z+1+d-\theta -\frac{2\theta }{d}}}=0.  \label{floweq2}
\end{equation}
Note that the new horizon and cutoff surface are located at $u_{h}=1$ and $\epsilon =r_{h}/r_{\epsilon }$. In addition, the Hawking temperature can be calculated as
\begin{equation}
T=-\frac{r_{h}^{z}f^{\prime }(u_{h})}{4\pi }.  \label{TH}
\end{equation}
Combining (\ref{floweq2}) and (\ref{TH}), we identify a residual scaling symmetry\footnote{This scaling symmetry was previously overlooked as conventional holographic modeling fixes the asymptotic behavior of $f(z)$, thereby breaking the symmetry.}
\begin{equation}
f\rightarrow f\lambda ,\quad
\bar{\omega}\rightarrow \bar{\omega}\lambda ,\quad 
\bar{T}\rightarrow \bar{T}\lambda ,\quad
\bar{m}^{2}\rightarrow \bar{m}^{2}\lambda,
\label{RS}
\end{equation}
where we have defined the rescaled temperature $\bar{T}\equiv T/r_{h}^{z}$. This novel symmetry will be exploited to reduce the number of free parameters.

\section{Machine learning algorithm}\label{ML}

We will recast the radial flow equation as a Neural ODE and then determine the metric function by minimizing an appropriate loss function.

\subsection{Neural ODE}

Neural ODE is an elegant combination of two important modeling methods: neural networks and ODEs \cite{Chen:2018wjc}. It provides a natural representation of continuous-time processes with high parameter efficiency. In Neural ODEs, the derivative of the hidden state $x$ with respect to the time $t$ is parameterized using a neural network $y$:
\begin{equation}
\frac{dx(t)}{dt}=y(x(t),t,\Omega ),  \label{NODE}
\end{equation}
where $\Omega $ represents the trainable parameters. Neural ODEs reduce memory usage during training through adjoint sensitivity methods for backpropagation. Consider a loss function
\begin{equation}
L\left( x( t_{1}) \right) =L\left(x( t_{0})
+\int_{t_{0}}^{t_{1}}y(x(t),t,\Omega )dt\right),  \label{Lxt}
\end{equation}
which depends on the solution of (\ref{NODE}). The adjoint sensitivity method introduces the adjoint state $a(t)=\partial
L/\partial x(t) $ to track the sensitivity of the loss with respect to the hidden state. Importantly, it has been proven that the adjoint state satisfies its own ODE  \cite{Chen:2018wjc}
\begin{equation}
\frac{da(t)}{dt}=-a(t)\frac{\partial y}{\partial x}.  \label{adjoint_eq}
\end{equation}
To calculate the gradients efficiently, one first solves the forward ODE (\ref{NODE}) to obtain $x(t) $. Then, starting from the final time, the adjoint equation (\ref{adjoint_eq}) is integrated backward in time to compute $a(t) $. During this backward integration, the gradient is calculated as
\begin{equation}\label{Omega_gradient}
\frac{\partial L}{\partial \Omega }=\int_{t_{1}}^{t_{0}}a(t) \frac{\partial y}{\partial \Omega }dt
\end{equation}
utilizing the chain rule and the adjoint state.

Now we convert the radial flow equation (\ref{floweq2}) into the Neural ODE. We first split the complex equation into real and imaginary parts:
\begin{eqnarray}
\frac{d\bar{\chi}_{\mathrm{re}}}{du} &=&\frac{2\bar{\omega}u^{z-1+d-\theta }}{f}\bar{\chi}_{\mathrm{re}}\bar{\chi}_{\mathrm{im}}, \\
\frac{d\bar{\chi}_{\mathrm{im}}}{du} &=&\frac{\bar{\omega}}{f}u^{z-1}\left[ u^{d-\theta }\left( \bar{\chi}_{\mathrm{im}}^{2}-\bar{\chi}_{\mathrm{re}}^{2}\right) +\frac{1}{u^{d-\theta }}\right] -\frac{1}{\bar{\omega}}\frac{m^{2}}{u^{z+1+d-\theta -2\theta /d}}.
\end{eqnarray}
Then we equate the radial coordinate $u$ to time $t$ and the response function $\bar{\chi}$ to state $x$. Furthermore, we will represent the metric function $f(u)$ through a fully connected neural network. All parameters in the neural-network representation of $f(u)$, together with the undetermined parameters in the flow equations, are optimized using the gradient in \eqref{Omega_gradient}.

\subsection{Inductive bias and loss function}

We will analyze the free parameters of our model and propose some inductive biases.

First, suppose that our field theory is dual to a black hole with a horizon. Consider the horizon located at $u_{h}=1$ and the black hole with the rescaled temperature $\bar{T}$. Then we can set the ansatz
\begin{equation}
f(u)=(1-u)\left[ 4\pi \bar{T}+(1-u)n(u,\gamma )\right] ,  \label{fansatz}
\end{equation}
where $n$ denotes a neural network with the parameters $\gamma $.

Second, in terms of $\bar{T}$, the bulk quantity $\bar{\omega}$ can find its field-theory correspondence
\begin{equation}
\bar{\omega}\equiv \frac{\omega }{r_{h}^{z}}=\frac{\omega }{T}\bar{T}.
\end{equation}
Due to the residual scaling symmetry (\ref{RS}), $\bar{T}$ (or $\bar{m}^{2}$) can be chosen arbitrarily, while $\bar{m}^{2}$ (or $\bar{T}$) should be
determined by fitting the data.

Third, we will assume $h(\omega )$ to be real\footnote{For the three holographic models in Section 5, one can prove that $h(\omega )$ is real by holographic renormalization \cite{Skenderis:2002wp, Papadimitriou:2016yit, Chemissany:2014xsa}. For the SYK model, the validity of the assumption can be judged by the minimum loss that we can achieve. }. As a result, in standard (alternative) quantization, $h(\omega )$ does not influence the imaginary part of $G_{s}$ ($G_{a}^{-1}$).

Fourth, the renormalization factor $\zeta$ does not depend on $\omega$, so it does not appear in the ratio of two Green functions at different frequencies. 

In terms of the above analysis about $h(\omega )$ and $\zeta $, we define the normalized field-theory Green functions 
\begin{equation}
\bar{G}_{a}\equiv \frac{\omega _{0}}{\omega }\frac{\mathrm{Im}G_{a}\left(
\omega \right) ^{-1}}{\mathrm{Im}G_{a}\left( \omega _{0}\right) ^{-1}},\quad 
\bar{G}_{s}\equiv \frac{\omega _{0}}{\omega }\frac{\mathrm{Im}G_{s}\left(
\omega \right) }{\mathrm{Im}G_{s}\left( \omega _{0}\right) },\label{Gbar}
\end{equation}
where $\omega _{0}$ is a reference frequency. Using (\ref{Ga}) and (\ref{Gs}), their bulk duals have the same form
\begin{equation}
\bar{G}=\frac{\mathrm{Re}\bar{\chi}\left( \omega ,\epsilon
\right) }{\mathrm{Re}\bar{\chi}\left( \omega _{0},\epsilon \right) }.
\end{equation}
The imaginary part of the retarded Green function $\mathrm{Im} G_s(\omega)$ could be identified as the spectral density of the operator $O(k=0)$. This ratio $\bar{G}$ will be understood as the output of Neural ODEs in holography.

Consider the situation that one can measure the Green function or spectral density of a field theory system either theoretically or experimentally within a frequency domain. One can construct the same ratios $\bar{G}_\text{data}$ according to \eqref{Gbar} from the measurement data. Now we can use these data to train our holographic model, such that the difference between the holographic output $\bar{G}$ and the field-theory data $\bar{G}_\text{data}$ is minimized. We choose the loss function to be
\begin{equation}
L=\sum_{\omega }\frac{1}{2}\left( \left\vert \frac{\bar{G}}{\bar{G}_\text{data}}
\right\vert -1-\ln \left\vert \frac{\bar{G}}{\bar{G}_\text{data}}\right\vert
\right), \label{loss}
\end{equation}
which is a combination of relative error and logarithmic difference. In App.~\ref{KL}, we give the field-theory motivation of this loss function. The loss function has the similar form as the Kullback–Leibler divergence between probability distributions of source $J$ determined by the partition function \eqref{pfft} in field theory.

\section{Examples}\label{epl}

Given the inductive biases in the last section, our model $\bar{G}$ still has the following free parameters
\begin{equation}
\gamma ,\;\epsilon ,\;\bar{T}\text{ (or }\bar{m}^{2}\text{)},\;z,\text{\ } \theta.  \label{free}
\end{equation}
We find it challenging to train all the parameters together. Therefore, we focus on training only $\gamma $ for simplicity. The other parameters are considered hyperparameters or are assumed to be fixed. The training scheme and hyperparameter tuning are detailed in App.~\ref{AD}.

\subsection{Conformal field theory}

We start from the CFT$_{1}$ at finite temperature. Using the conformal symmetry, one can derive the Green function of a scalar operator with scaling dimension $\delta $
\begin{equation}
G_{\mathrm{CFT}}\sim \frac{\Gamma \left( \delta -\frac{i\omega }{2\pi T}
\right) }{\Gamma \left( 1-\delta -\frac{i\omega }{2\pi T}\right) },
\label{GCFT}
\end{equation}
where we have adapted the alternative quantization and \textquotedblleft $\sim $\textquotedblright\ indicates that we are ignoring the finite factor that is real and
independent of $\omega $. This Green function can also be obtained in terms of HWRG.

The two-dimensional $(d=0)$ AdS black hole is described by the metric in \eqref{f2D} with the function
\begin{equation}
f(u)=1-u^{2}  \label{PAdS}
\end{equation}
or, equivalently, by the metric in \eqref{f12} with exponents $z=1$ and $\theta/d=0$. Upon inserting these exponents into (\ref{floweq2}), the flow equation can be solved analytically. Using (\ref{Ga}) and taking
\begin{equation}
\bar{m}^{2}=\delta\left( \delta -1\right),\quad r_{h}=2\pi T,\quad h=-\delta ,\quad \epsilon \rightarrow 0,
\end{equation}
one exactly recovers the Green function (\ref{GCFT}).

Let's generate the data for machine learning. We first normalize (\ref{GCFT}) to read the data $\bar{G}_\text{CFT}$. Next, we specify the scaling dimension $\delta =0.2$ and the reference frequency $\omega _{0}/(2\pi T)=0.01$. Then we sample $\omega /(2\pi T)$ uniformly between 0.01 and 5, which gives 500 data points\footnote{For all examples in this paper, we keep the data with the same frequency range.}.

\begin{figure}
\centering 
\includegraphics[width=.6\textwidth
]{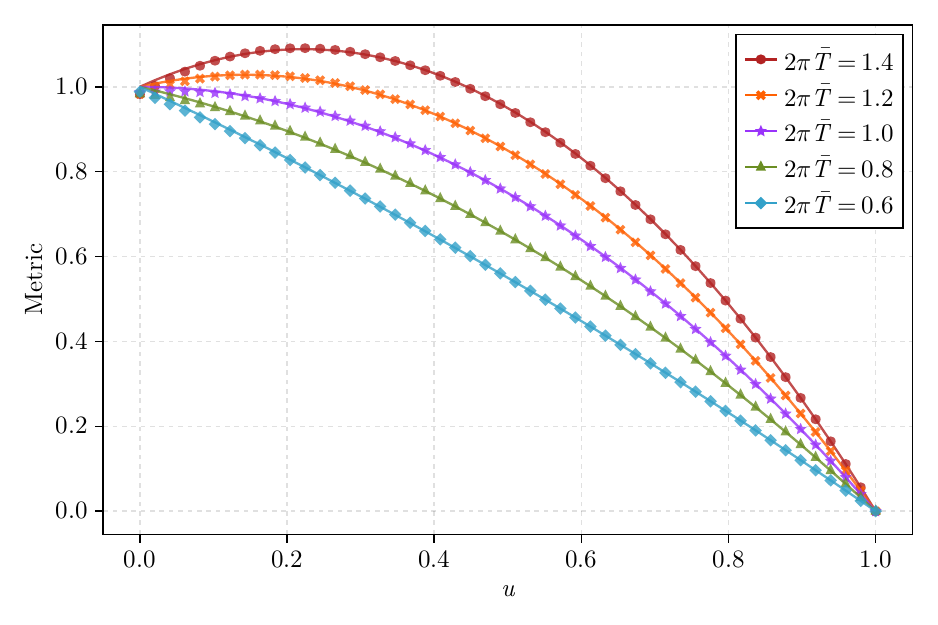}
\caption{The metric functions learned from CFT$_{1}$ at various $\bar{T}$. The solid and marker lines represent the AdS$_2$ black holes and the prediction of Neural ODEs, respectively.}
\label{RTCFT}
\end{figure}

In terms of the residual scaling symmetry (\ref{RS}), we select $\bar{m}^{2}=\delta
\left( \delta -1\right) =-0.16$ for convenience\footnote{
If one selects other values, the emergent metric will be the same after a
rescaling.}. Suppose the cutoff should approach zero but  $\bar{T}$ is unknown. Then we set $\epsilon=10^{-4}$ and treat $\bar{T}$ as a hyperparameter, scanning it over the range $2\pi \bar{T}\in \left[ 0.6,1.4 \right] $ in steps of 0.2. Interestingly, we find that their loss functions are very close at different $\bar{T}$ (see table \ref{LM} in App.~\ref{AD}) and the emergent metric meets the following form exactly (see Fig.~\ref{RTCFT})
\begin{equation}
 f(u)=(1-u)\left[ 1+u\left( 4\pi \bar{T}-1\right) \right] . \label{TAdS}
\end{equation}

The metric \eqref{f2D} with (\ref{TAdS}) describe an AdS$_{2}$ at finite temperature
\begin{equation}
ds^{2}=d\rho^2-\frac{4\pi^2}{\bar{\beta}^2}\sinh^2\rho d t^2,
\label{AdS R}
\end{equation}
with $\bar{\beta}=1/\bar{T}$, via the transformation of radial coordinate 
\begin{align}
\cosh\rho =1-\frac 1{ 2\pi\bar{T} }\left(1-\frac1u\right).\label{rtrf}
\end{align}
Meanwhile, the metric function (\ref{PAdS}) is a special case of \eqref{TAdS} with $\bar{T}=\frac{1}{2\pi}$, which describes a region contained within the Poincare patch. Put differently, the neural ODE has discovered a family of hidden spacetimes with parameter $\bar{T}$  from the CFT data.

\subsection{Lifshitz spacetime}

Lifshitz spacetime is a simple bulk geometry with asymptotic scaling symmetry, but it is not asymptotically AdS. The metric is given by \eqref{f12} and \eqref{f12f} with the function and parameters
\begin{equation}
f=1-u^{2},\quad d=2,\quad z=2,\quad \theta =0.
\end{equation}
We will regard the data as the Green function generated by Lifshitz spacetime.
In \cite{Balasubramanian:2009rx}, the analytical form of the Green function in the standard quantization has been derived:
\begin{equation}
G_{\mathrm{LIF}}\sim \frac{\Gamma \left[ \frac{1}{2}\left( 1-i\bar{\omega}+
\sqrt{4+\bar{m}^{2}}-\sqrt{1-\bar{\omega}^{2}}\right) \right] }{\Gamma \left[
\frac{1}{2}\left( 1-i\bar{\omega}-\sqrt{4+\bar{m}^{2}}-\sqrt{1-\bar{\omega}
^{2}}\right) \right] }\frac{\Gamma \left[ \frac{1}{2}\left( 1-i\bar{\omega}+
\sqrt{4+\bar{m}^{2}}+\sqrt{1-\bar{\omega}^{2}}\right) \right] }{\Gamma \left[
\frac{1}{2}\left( 1-i\bar{\omega}-\sqrt{4+\bar{m}^{2}}+\sqrt{1-\bar{\omega}
^{2}}\right) \right] },
\end{equation}
where the frequency has been measured in units of $r_{h}=2\pi T$.

\begin{figure}
\centering 
\raisebox{-0.5mm}{
\includegraphics[height=.32\textwidth
]{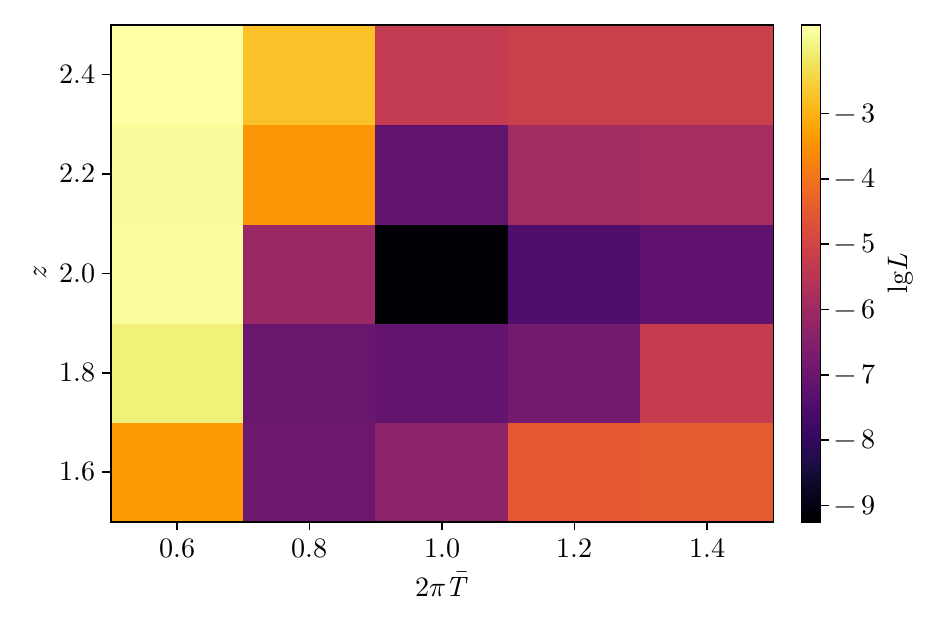}
}
 \hspace{0.0cm}
\includegraphics[height=.32\textwidth
]{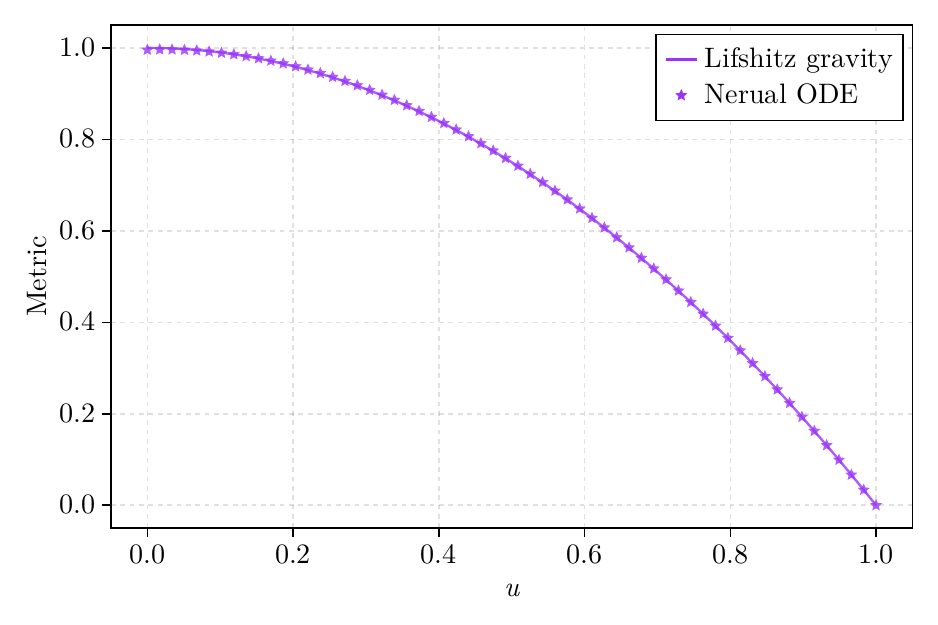}

\caption{(Left) Loss distribution as a function of hyperparameters $z$ and $\bar{T}$ for Lifshitz gravity. (Right) The metric function learned by Neural ODEs at optimal hyperparameters $(2\pi \bar{T},z)=(1,2)$.}
\label{HTLIF}
\end{figure}

Now we generate the normalized data $\bar{G}_\text{LIF}$, for which we set $ \bar{\omega}_{0}=0.01$ and $\bar{m}^{2}=-1$. We treat $\left( \bar{T} ,z\right) $ as two hyperparameters and select $\bar{m}^{2}=-1$ for convenience. The rest parameters $\left( \epsilon ,\theta \right) $ are
fixed as $\left( 10^{-4},0\right) $. We scan the hyperparameters in the ranges $2\pi \bar{T}\in \left[ 0.6,1.4\right] $ with a step of 0.2 and$\ z\in \left[ 1.6,2.4\right] $ with the same step. From the left panel of Fig.~\ref{HTLIF}, one can identify the optimal hyperparameters. Fixing $\left( \bar{T},z\right) $ as the optimal values, we conduct the training to learn the metric. The training result has been plotted in the right panel of Fig.~\ref{HTLIF}.

\subsection{Hyperscaling violation}

We further consider the data with hyperscaling violation \cite{Dong:2012se}. Suppose that the metric function and three parameters are
\begin{equation}\label{hybh}
f=1-u,\quad d=1,\quad z=1,\quad \theta =1.
\end{equation}
Equation (\ref{hybh}) allows us to obtain an analytical Green function. By solving the KG equation in the black hole background and applying the infalling boundary condition, we obtain a scalar field solution given by
\begin{equation}\label{solh}  \phi_c\sim  \Gamma(1-2i\bar{\omega})e^{\pi \bar{\omega}}I_{-2i\bar{\omega}}(2\bar{m}\sqrt{1-u}),
\end{equation}
where $I_n(x)$ is a modified Bessel function of the first kind. The
asymptotic behavior of the solution in (\ref{solh}) is
\begin{equation}  \phi_c\sim A+B u+\cdots.
\end{equation}
Thus, we obtain the Green function in the standard quantization
\begin{equation}
G_{\mathrm{HV}}\sim i\bar{\omega} -\frac{\bar{m}I_{1-2i\bar{\omega}}(2\bar{m})}{I_{-2i\bar{\omega}}(2\bar{m})}.
\end{equation}

\begin{figure}[ht]
\centering 
\raisebox{-0.5mm}{
\includegraphics[height=.32\textwidth
]{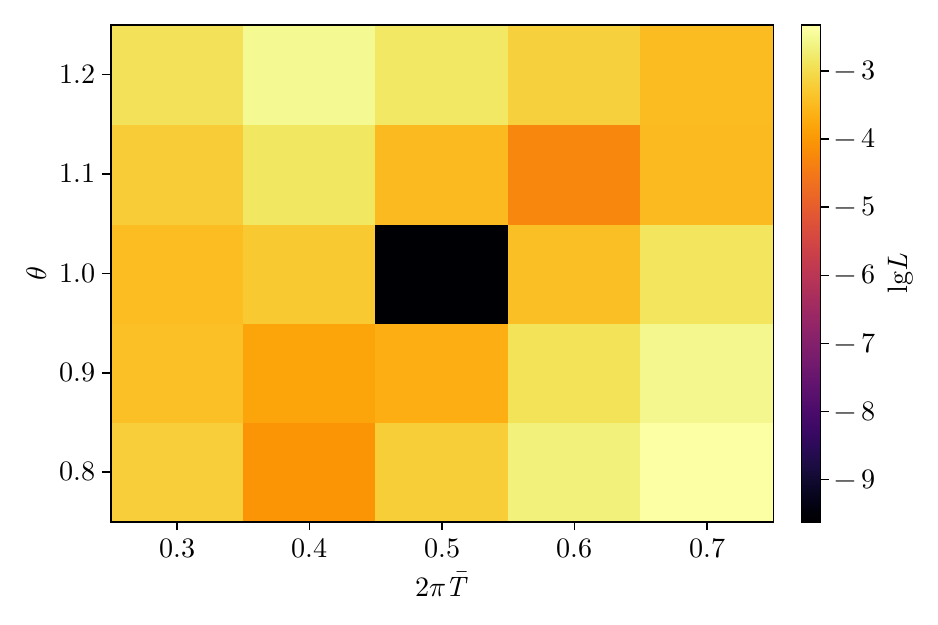}
}
\hspace{0.0cm}
\includegraphics[height=.32\textwidth
]{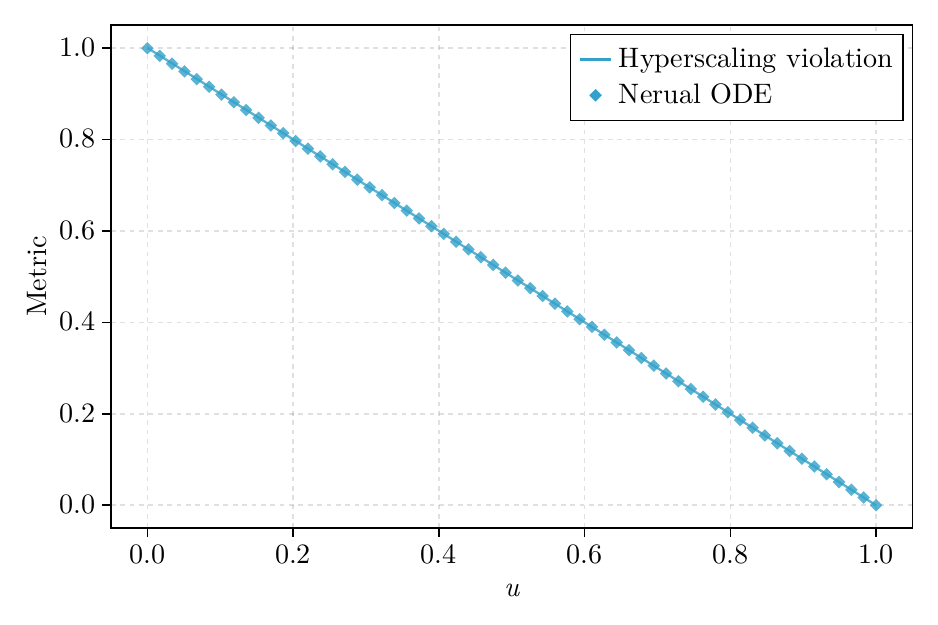}
\caption{(Left) Loss distribution as a function of hyperparameters $\theta$ and $\bar{T}$ for the theory with hyperscaling violation. (Right) The metric function learned by Neural ODEs at optimal values of hyperparameters $(2\pi\bar{T},\theta)=(0.5,1)$.}
\label{HTHV}
\end{figure}

We generate the normalized data $\bar G_\text{HV}$ with $\bar{\omega}_{0}=0.01$ and $\bar{m}^{2}=-1$. We fix $\left( \epsilon ,z\right) =\left(
10^{-4},1\right) $ and select $\bar{m}^{2}=-1$ for convenience. With these preparation, we scan the hyperparameters $\left( \bar{T},\theta \right) $ in the ranges $2\pi \bar{T}\in 
\left[ 0.3,0.7\right] $ and $\theta \in \left[ 0.8,1.2 \right] \;$ with the step size of 0.1. The results of hyperparameter tuning and metric learning can be found in Fig.~\ref{HTHV}.

\subsection{The Sachdev-Ye-Kitaev model}
The SYK model is a quantum mechanical model of $N$ Majorana fermions $\{\chi_j|j=1,2,\cdots,N\}$ with random $q$-body interactions \cite{Kitaev:1,Kitaev:2,Maldacena:2016hyu}. The SYK model is analytically tractable in the large-$N$ limit.  At low energies it develops an emergent conformal symmetry, while at high energies it crosses over to weakly-coupled fermion behavior.  The infrared dynamics is captured by Jackiw–Teitelboim (JT) gravity on AdS$_2$ coupled to matter fields \cite{Almheiri:2014cka,Maldacena:2016upp,Engelsoy:2016xyb,Gross:2017hcz,Goel:2023svz}, whereas the ultraviolet regime is expected to deviate from AdS geometry. These contrasting limits make the SYK model an ideal testing ground for our learning framework.

The SYK Hamiltonian is given by
\begin{equation}
H_\text{SYK}=i^{q/2}\sum_{1\leq j_1<\cdots<j_q\leq N}J_{j_1,...,j_q}\chi_{j_1}\ldots\chi_{j_q},
\end{equation}
where these Majorana fermions obey $\{\chi_j,\chi_k\}=\delta_{jk}$. The coupling constants $J_{j_{1},\cdots j_{q}}$ are Gaussian random real variables with mean zero and variance
\begin{equation}
\overline{J_{j_1,\cdots,j_q}^2}=\frac{2^{q-1}}q\frac{\mathcal{J}^2(q-1)!}{N^{q-1}},
\end{equation}
where the tunable parameter $\mathcal{J}$ dictates the characteristic energy scale in the large $q$ limit.

At finite temperature and finite $q$, the two-point function is obtained by numerically solving the Schwinger–Dyson equations.  In the large-$q$ limit, $1 \ll q^2 \ll N$, an analytic expression valid at all temperatures is available \cite{Maldacena:2016hyu}:
\begin{align}
G(\tau)& 
= \frac1N\sum_{j=1}^N\langle \mathcal  T\chi_j(\tau)\chi_j\rangle_\beta  
=\frac12\left[\frac{\cos(\pi v/2)}{\cos\left(\pi v\left(\frac12-\frac{|\tau|}\beta\right)\right)}\right]^{2/q}\label{sykG}, \quad 0<\tau<\beta,
\end{align}
with anti-periodicity $G(\tau+\beta)=-G(\tau)$, where the dimensionless parameter $v\in(0,1)$ is obtained by solving
\begin{equation}
    \beta\mathcal{J} =\frac{\pi v}{\cos(\pi v/2)}.
\end{equation}

In this paper, we focus on the Green function of the composed bosonic operators
\begin{equation}
O_{jk}=i\chi_j\chi_k,
\end{equation}
whose scaling dimensions is twist of that of single Majorana fermions
\begin{equation}
\left[ O_{jk}\right]_\text{IR}=\delta=2/q,\quad \left[ O_{jk}\right]_\text{UV}=0.
\end{equation}
where ``IR'' refers to the conformal fixed point and ``UV'' refers to the free fermion fixed point.
The Green function at the leading $1/N$ expansion is
\begin{equation}
  G_{\mathrm{SYK}}(\tau)=\dfrac{1}{N^2}\sum_{jk}^N\langle O_{jk}(\tau )O_{jk}\rangle_{\beta}=G(\tau)^2.
\end{equation}
Taking the Fourier transform $\tau\to \omega_n$ and Wick rotation $\omega_n\to i \omega+0^+$, we can get the retarted Green function in frequency region
\begin{align}\label{GSYK}
    	G_{\mathrm{SYK}}(\omega)=\frac{i\mathcal{A}}{8\pi(n q+2 v)}
	\left[-e^{\frac{i 2\pi  v}{q}} \, _2F_1\left(\frac{4}{q},\frac{2}{q}+\frac{n}{v};\frac{n}{v}+\frac{2}{q}+1;-e^{i \pi  v}\right)\right. \notag \\ 
	\left. +e^{-\frac{i 2\pi  v}{q}}\, _2F_1\left(\frac{4}{q},\frac{2}{q}+\frac{n}{v};\frac{n}{v}+\frac{2}{q}+1;-e^{-i \pi  v}\right)\right],
\end{align}
where constant $\mathcal{A}=qT\left(2\cos\frac{\pi v}{2}\right)^{\frac{4}{q}}$ and $n=\frac{i\omega}{2\pi T}+0^+$. As $v$ approaches the upper bound ($ v \to 1 $), the SYK system converges to a conformal regime. Conversely, as $v$ approaches the lower bound ($v \to 0$), the system approaches a UV-free theory.

We construct the normalized data $\bar{G}_{\mathrm{SYK}}$ for several choices of $v$ near $1$ at fixed $\omega_{0}/(2\pi T)=0.01$. For comparison with $\mathrm{CFT}_{1}$, we take $\bar{m}^{2}=-0.16$ (i.e., $q=10$) and scan the hyperparameters $(\epsilon,\bar{T})$. After hyperparameter optimization, we train the model and report the following results.

\begin{figure}[ht]
\centering 
\includegraphics[width=.6\textwidth
]{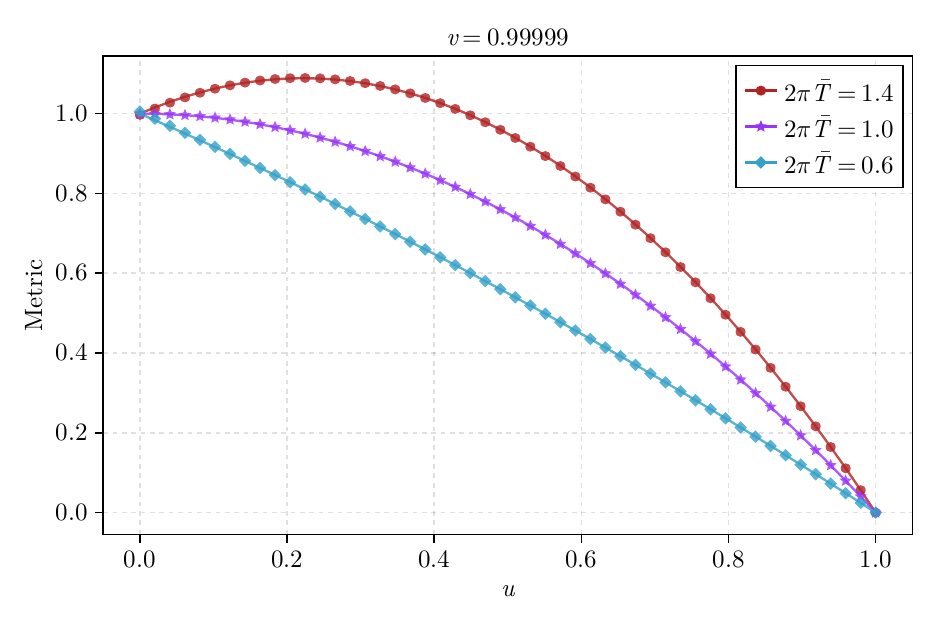}
\caption{Comparison between AdS$_2$ black hole metrics (solid lines) and learned metrics (marker lines) at different $\bar{T}$ values for the SYK data near the conformal limit. The cutoff is fixed as $\epsilon=10^{-4}$.}
\label{RTSYK99999}
\end{figure}

As $v\to 1$, the SYK model is captured by the AdS$_2$ black hole (Fig.~\ref{RTSYK99999}), in agreement with $\mathrm{CFT}_1$ expectations.

\begin{figure}
\centering 
\includegraphics[width=.49\textwidth
]{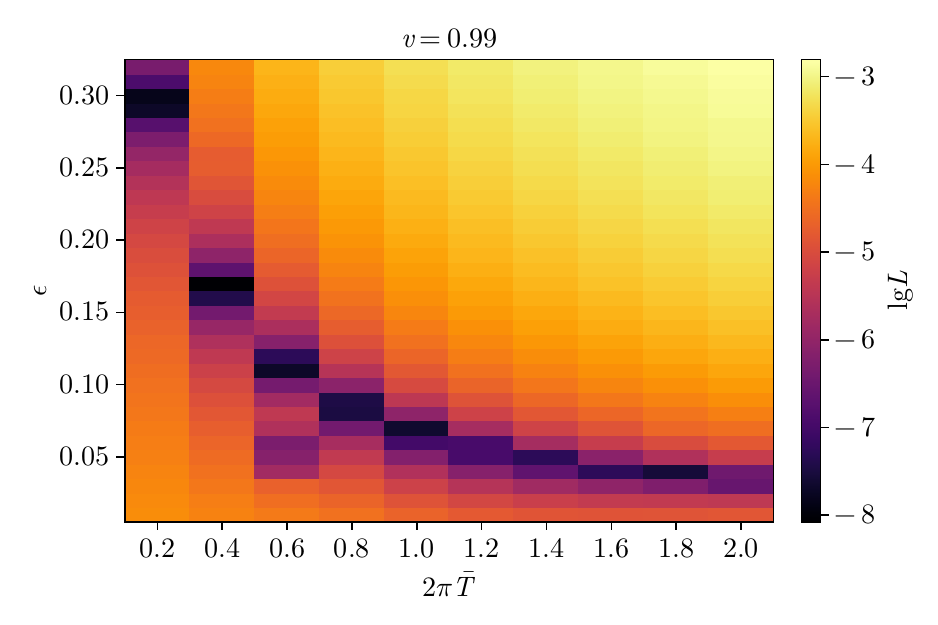}
\hspace{0.0cm}
\includegraphics[width=.49\textwidth
]{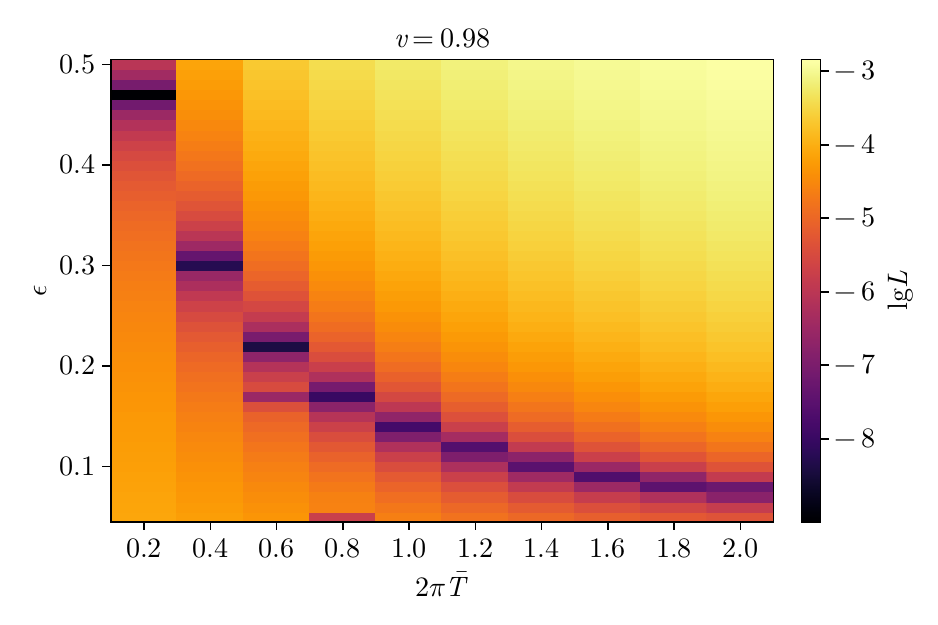}
\hspace{0.0cm}
\includegraphics[width=.49\textwidth
]{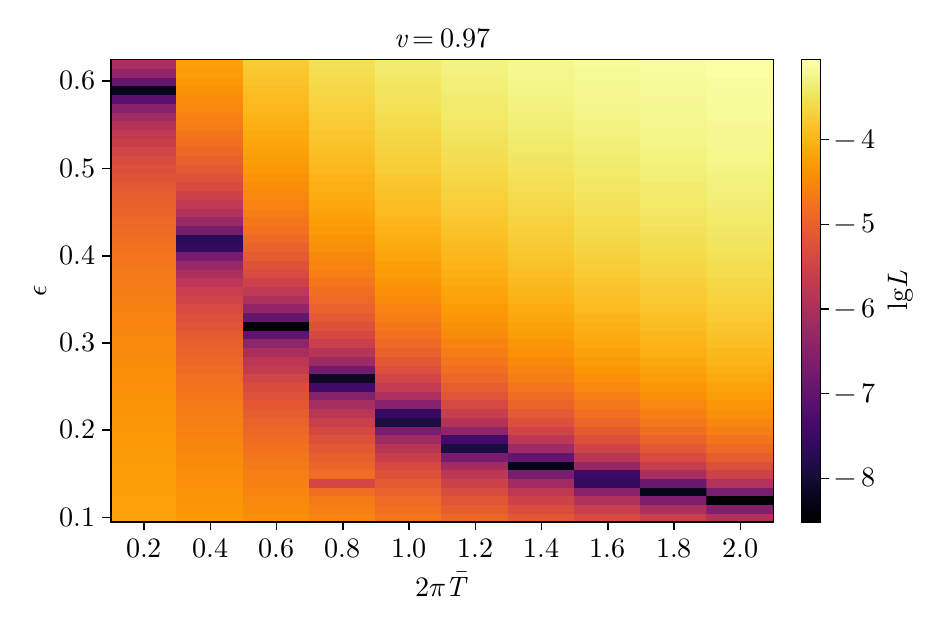}
\hspace{0.0cm}
\includegraphics[width=.49\textwidth
]{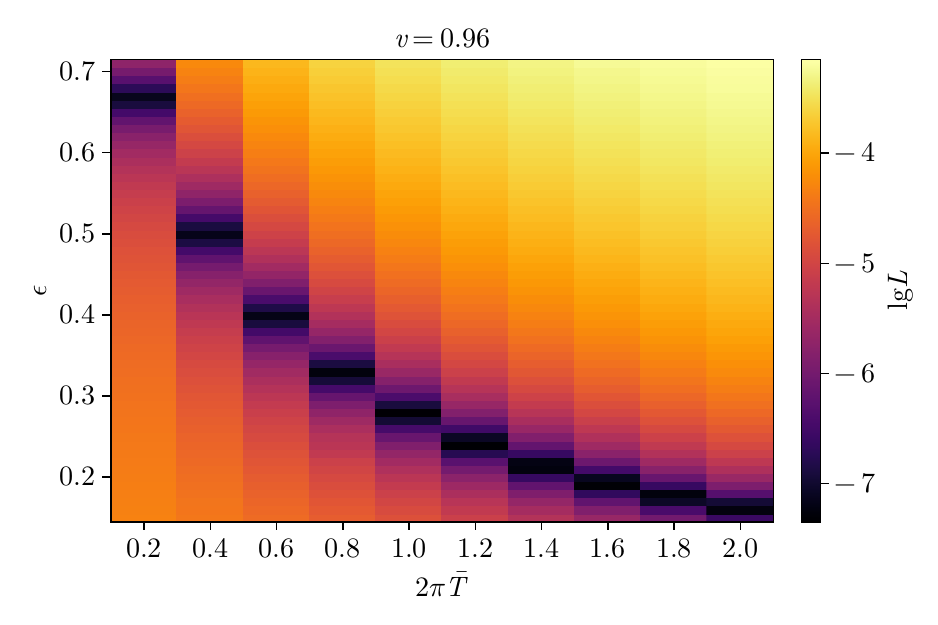} 
\caption{Loss distributions as functions of the hyperparameters $\epsilon$ and $\bar{T}$ for SYK model at various $v$. In each panel, the darkest squares form a monotonous curve.}
\label{HTSYK}
\end{figure}

\begin{figure}
\centering 
\includegraphics[width=.6\textwidth
]{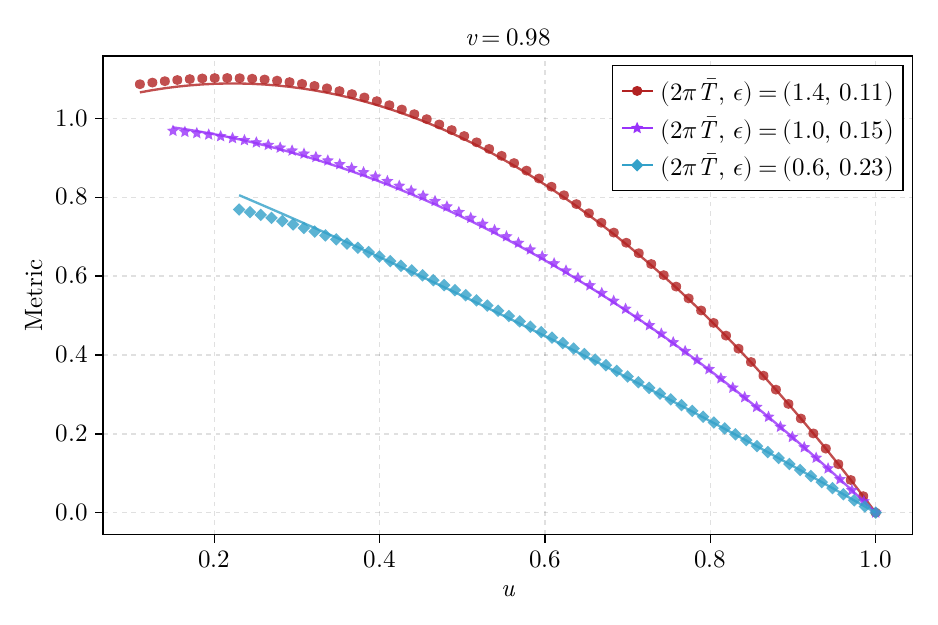}
\caption{Neural ODE-derived metrics for SYK model slightly deviating from the conformal limit, with $(\epsilon, \bar{T})$ fixed on the critical curve. Solid and marker lines represent AdS$_2$ black hole and learned metrics respectively.}
\label{RTSYK98}
\end{figure}

\begin{figure}
\centering 
\includegraphics[width=.9\textwidth
]{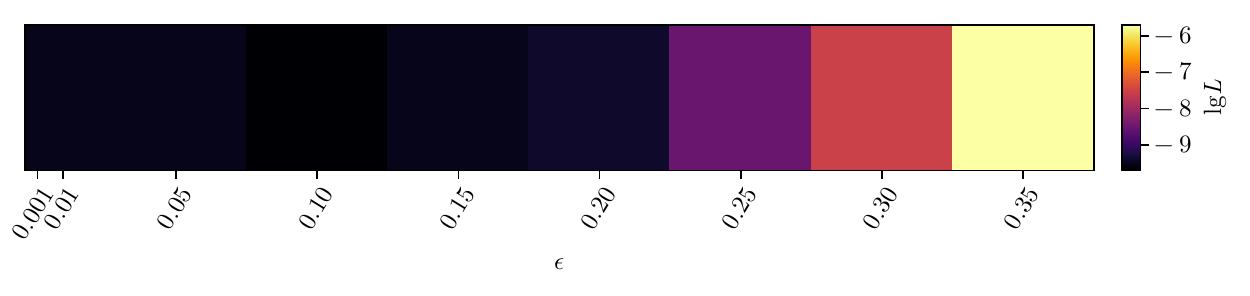}
\caption{Loss distributions as values of $\epsilon$ for SYK model at $v=0.98$ and $2\pi \bar{T}=1$.}
\label{HTSYK.98below}
\end{figure}

\begin{figure}
\centering 
\includegraphics[width=.49\textwidth
]{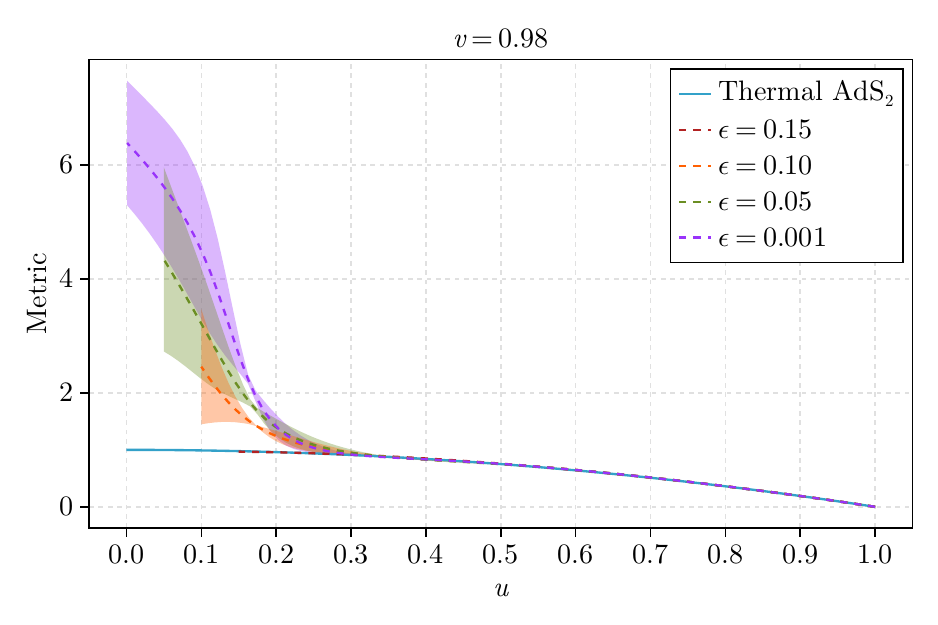}
\includegraphics[width=.49\textwidth
]{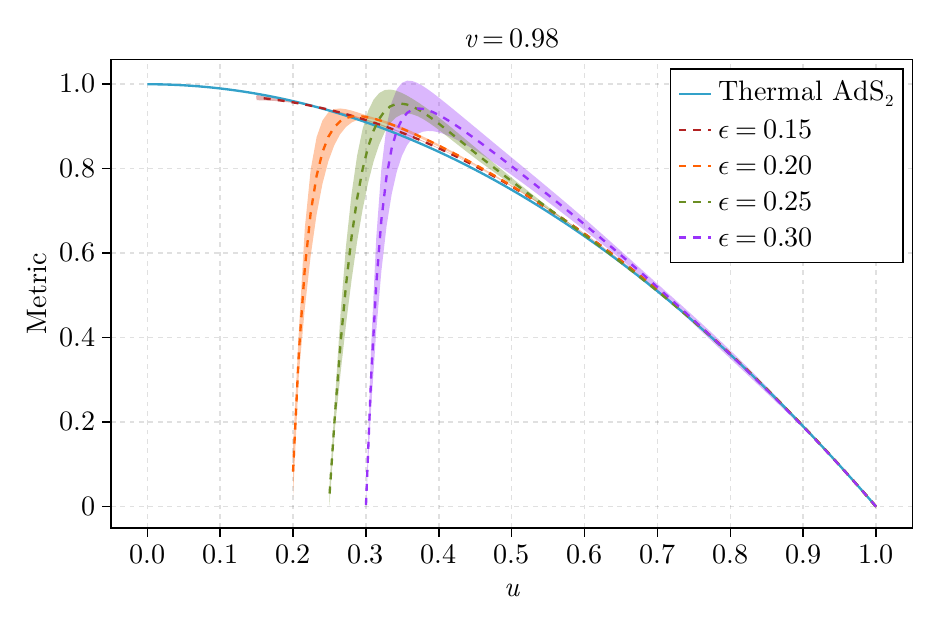}
\caption{Comparison of learned metrics with different $\epsilon$ at fixed $2\pi \bar{T}=1$. Shaded regions represent \text{1$\sigma$} confidence intervals, plotted using five training results with similar losses but different initial parameters.}
\label{RTSYK98below}
\end{figure}

For $v$ slightly below $1$, hyperparameter tuning reveals a critical curve in the $(\epsilon,\bar{T})$ plane (Fig.~\ref{HTSYK}). Along this curve, the learned metric closely matches an AdS$_2$ black hole with a finite cutoff (Fig.~\ref{RTSYK98})
\footnote{For Fig.~\ref{RTSYK98} we shift the critical cutoff to $\epsilon=\epsilon_{0}+0.01$, where $\epsilon_{0}$ denotes the darkest squares in Fig.~\ref{HTSYK}. The difference is negligible; hence we do not distinguish $\epsilon$ from $\epsilon_{0}$ elsewhere.}. Above the curve, the loss (with sufficient training) transitions from slow to rapid growth\footnote{This loss differs from that shown in the heatmap of Fig.~\ref{HTSYK}, which is based on fast but insufficient training; see App.~\ref{AD}.}, whereas below it the loss remains small and nearly constant. Far from the curve (on either side), the learned metrics deviate strongly from AdS$_2$; moreover, the metric below the curve is highly sensitive to initialization, while that above is much less so. Strikingly, along the critical curve the learned metric is insensitive to initialization. These trends (Figs.~\ref{HTSYK.98below} and \ref{RTSYK98below}) suggest that the data are most compatible with gravitational degrees of freedom along the critical curve.

\begin{figure}
\centering 
\includegraphics[height=.33\textwidth
]{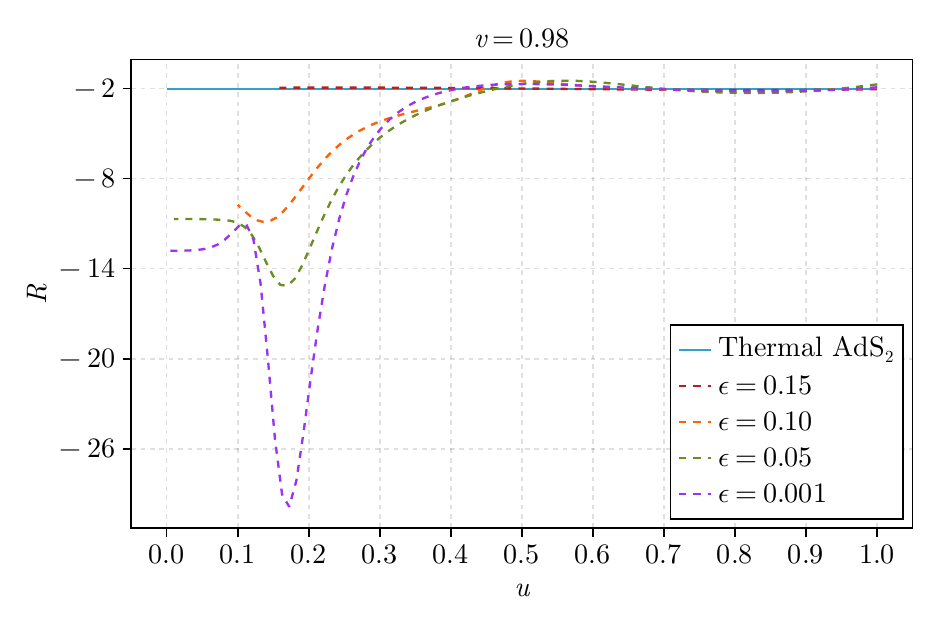}
\includegraphics[height=.33\textwidth
]{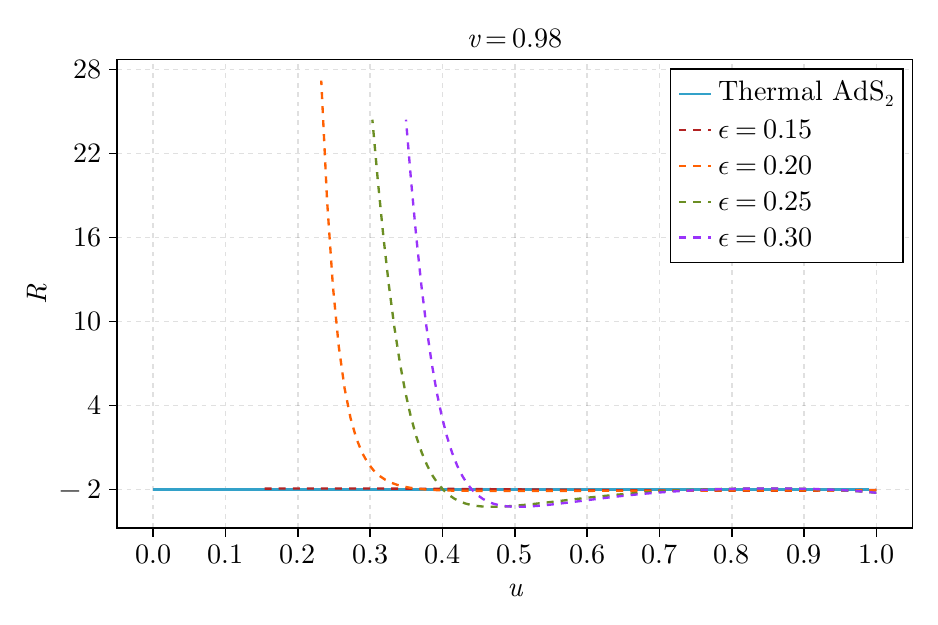}
\caption{The Ricci scalar of learned metrics with different $\epsilon$ at $2\pi \bar{T}=1$.}
\label{curvature}
\end{figure}

Motivated by Ref.~\cite{Nebabu:2023iox}, we also plot the curvature of the learned metric (Fig.~\ref{curvature}). Above the critical curve, the Ricci scalar grows rapidly in the UV and tends toward divergence as the cutoff increases, and it remains positive-opposite to the sign reported in Ref.~\cite{Nebabu:2023iox}. Below the curve, a region of negative curvature appears, whose magnitude increases as $u$ decreases, but this growth does not reach the cutoff.

\begin{figure}[ht]
\centering 
\raisebox{-0.5mm}{
\includegraphics[height=.32\textwidth
]{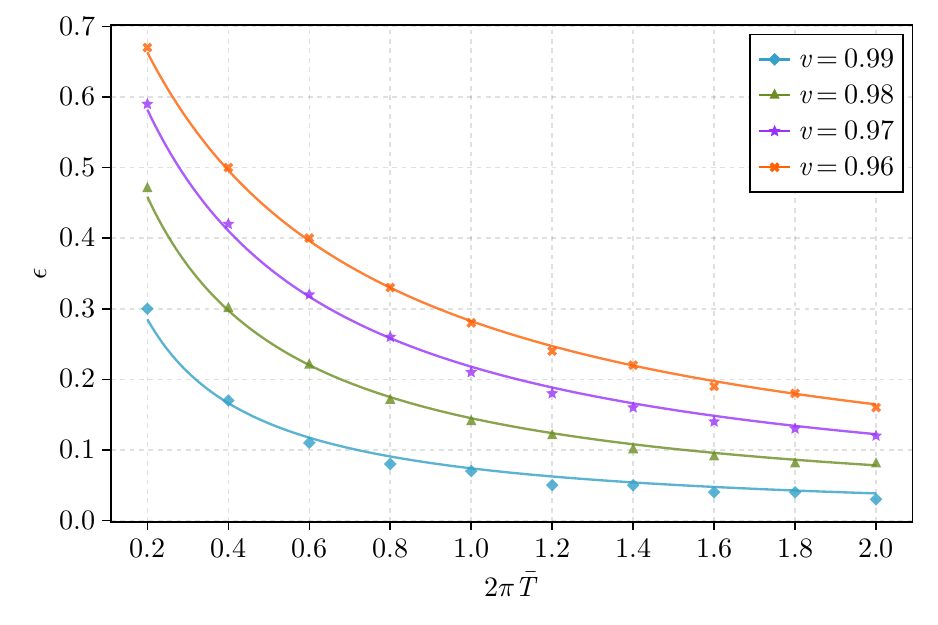}
}
\hspace{0.0cm}
\includegraphics[height=.32\textwidth
]{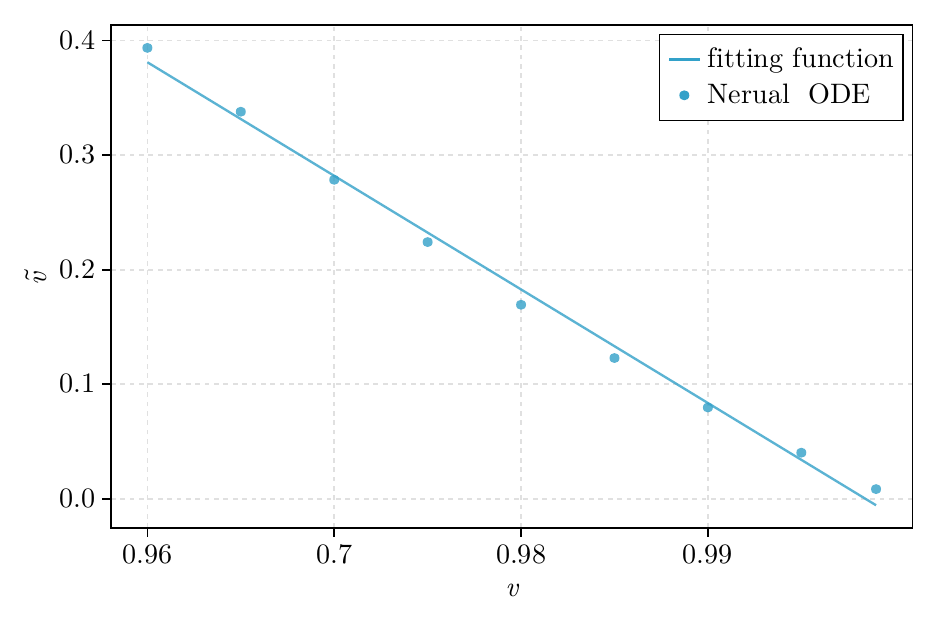}
\caption{(Left) Comparison of critical curves predicted by Neural ODEs (marker lines) with the analytical formula \eqref{ub_T_vt} (solid lines). (Right) The relationship between $v$ and $\tilde{v}$ is obtained by linear fitting.}
\label{mbb}
\end{figure}

We now account for the critical curves on the $\epsilon-T$ plane in Fig.~\ref{HTSYK}. If the bulk geometry is close to an AdS$_2$ black hole with a finite cutoff, the associated Green function can be written as
\begin{equation}
G_{a}\sim \frac{H_{1}\Gamma _{1}-H_{2}\Gamma _{2}}{A_{1}\Gamma
_{1}-A_{2}\Gamma _{2}}  \label{GaCFT},
\end{equation}where
\begin{eqnarray}\label{GaCFTwhere}
\Gamma _{1} &=&\left( -2\right) ^{2\delta }\tilde{v}\Gamma \left( 1-\delta -
\frac{i\bar{\omega}}{2\pi\bar{T}}\right) \Gamma \left( \frac{1}{2}+\delta \right) ,\;\Gamma
_{2}=\;2\left( -\tilde{v}\right) ^{2\delta }\Gamma \left( \frac{3}{2}-\delta
\right) \Gamma \left( \delta -\frac{i\bar{\omega}}{2\pi \bar{T}}\right) ,  \notag \\
H_{1} &=&_{2}F_{1}(1-\delta ,1-\delta -\frac{i\bar{\omega}}{2\pi \bar{T}};2-2\delta ;-2\tilde{v}
),\;H_{2}=\;_{2}F_{1}(\delta ,\delta -\frac{i\bar{\omega}}{2\pi \bar{T}};2\delta ;-2\tilde{v}), 
\notag \\
H_{3} &=&_{2}F_{1}(2-\delta ,1-\delta -\frac{i\bar{\omega}}{2\pi \bar{T}};2-2\delta ;-2\tilde{v}
),\;H_{4}=\;_{2}F_{1}(1+\delta ,\delta -\frac{i\bar{\omega}}{2\pi \bar{T}};2\delta ;-2\tilde{v}), \notag \\
A_{1} &=&hH_{1}\sqrt{1+2\tilde{v}}-iH_{1}\tilde{v}\frac{\bar{\omega}}{2\pi \bar{T}}-H_{3}\left(
1+2\tilde{v}\right) \left( \delta -1\right) ,  \notag \\
A_{2} &=&hH_{2}\sqrt{1+2\tilde{v}}-iH_{2}\tilde{v}\frac{\bar{\omega}}{2\pi \bar{T}}+H_{4}\left(
1+2\tilde{v}\right) \delta , \ \ \ \ \;
\tilde{v}=\frac{2\pi \bar{T}\epsilon}{1-\epsilon} .
\end{eqnarray}
Suppose that the data $\bar{G}_\text{SYK}$ generated by (\ref{GSYK}) can be approximately described by (\ref{GaCFT}); then it should depend on $\omega$ and $T$ only through the single scaling variable $\omega/T$. From the last equation in \eqref{GaCFTwhere}, we obtain a relation between $\epsilon$ and $T$,
\begin{equation}
\epsilon = \frac{1}{1+2\pi \bar{T}/\tilde{v}}, \label{ub_T_vt}
\end{equation}
with a parameter $\tilde{v}$ depending only on $v$. Relation (\ref{ub_T_vt}) fits the $(\epsilon,\bar{T})$ critical curves predicted by Neural ODEs, and the fitting parameter $\tilde{v}$ is found to depend linearly on $v$ via $\tilde{v}=9.9(1-v)$, as shown in Fig.~\ref{mbb}. Thus, dual to the parameter $v$ in the large-$q$ SYK, which characterizes the deviation from CFT$_1$, the Neural-ODE parameter $\tilde{v}$ likewise characterizes the deviation from AdS$_2$ asymptotics.

\section{Discussion}\label{Disc}
In this work, we proposed a novel approach to uncovering bulk geometries beyond traditional asymptotically AdS spacetimes, utilizing the HWRG combined with machine learning techniques. Our framework allows Neural ODEs to learn the finite UV cutoffs and various asymptotic geometries, such as those with Lifshitz and hyperscaling-violated behaviors.

Notably, this method effectively describes the geometric bulk of the SYK model slightly deviating from the conformal limit. A key observation is the presence of a critical curve in the $\epsilon-\bar{T} $ plane, associated with the AdS$_2$ black hole. This suggests that the SYK model which slightly deviates from the conformal limit can still be described by the AdS$_2$ black hole, but with the cutoff shifted towards the IR. This is reminiscent of $T\bar{T}$ deformations and their holographic implications  \cite{McGough:2016lol, Gross:2019uxi, Gross:2019ach, Hartman:2018tkw, Kraus:2018xrn}. Furthermore, the inferred linear relationship between the SYK coupling $v$ and the bulk parameter $\tilde{ v}$ effectively establishes a holographic dual.

Although our work supports the conventional view that the SYK model is associated with a well-defined geometry below a cutoff, we observe a trend of increasing loss and curvature divergence as the cutoff moves away from the critical curve and towards the horizon. The connection between these observations and ref.~\cite{Nebabu:2023iox} remains unclear and warrants further investigation.

To further extend or apply out framework of holographic machine learning, we suggest the following future research directions:
\begin{itemize}
\item \textbf{Spatial Dependence:} This paper primarily focuses on Green functions at zero wavevector. However, our framework can be directly extended to finite wavevectors. Accessing spacelike correlator information should enable the machine to better constrain the possible geometries and parameters.
\item \textbf{Fermion:} This paper uses Green functions of scalar operators. Future research can consider using fermionic Green functions as data, for instance, data from angle-resolved photoemission spectroscopy (ARPES). This would aid in exploring the bulk physics near the Fermi surface and in deeply understanding the holographic duals of strongly correlated electron systems.
\item \textbf{Entanglement:} Growing evidence suggests that spacetime may emerge from quantum entanglement \cite{Maldacena:2001kr,Ryu:2006bv,VanRaamsdonk:2010pw,Swingle:2009bg,Maldacena:2013xja}.  A promising direction for future work is to compare the similarities and differences between reconstructing spacetime metrics from entanglement entropies and from correlation functions.

\item \textbf{Wormhole:} Wormhole geometries in holographic duals are among the most non-trivial manifestations of quantum entanglement \cite{Maldacena:2013xja, Maldacena:2001kr, Maldacena:2004rf}. Future research could attempt to reconstruct wormhole geometries using machine learning. This can be achieved by training neural networks to learn the bulk geometry from correlation functions in thermal-field double states \cite{Maldacena:2018lmt, Cottrell:2018ash, Maldacena:2017axo}. We anticipate that machine-learning techniques built on boundary correlation functions can reconstruct the wormhole geometry even at late times, when entanglement-entropy data alone are insufficient \cite{Susskind:2014moa}.
\item \textbf{Higher Frequency:} This paper only studied the lower-frequency Green functions. Future research can explore situations with high frequency. In HWRG, the effects of high-energy modes are integrated out and embodied in the high-order terms in the low-energy effective action. Therefore, we must consider the coupling $h$ of the double-trace operators that we have ignored.

\item \textbf{Learning real data.} We anticipate that the present approach can be applied to real-world data from experiments, such as those on QCD and strange metals. These strongly coupled systems have not been adequately understood by existing theories. Integrating holography with machine learning within a sufficiently universal framework could yield more effective models for these phenomena.

\end{itemize}

\acknowledgments
We thank Souvik Banerjee, Xian-Hui Ge, Yan Liu, and Yu Tian for their helpful discussions.
SFW was supported by NSFC grants (No.12275166 and No.12311540141). ZYX acknowledges funding from the DFG through the Collaborative Research Center SFB 1170 “ToCoTronics” (Project ID 258499086 – SFB 1170), support from the National Natural Science Foundation of China under Grant No. 12075298, and support from the Berlin Quantum Initiative.

\appendix
\section{Derivation of the Green function in standard quantization}\label{stdc}
Building on the alternative quantization introduced in section \ref{HWRG}, we outline the essential steps to derive the two-point function in standard (Dirichlet) quantization.

Starting from the alternate quantization and the HWRG framework, the dictionary at finite radial cutoff is given by
\begin{align}\label{altZ}
    \langle e^{i\int JO} \rangle_{\Lambda}= \int_{r \geq r_{\epsilon}} D \phi \ e^{iS_0[\phi]+iS_B}.
\end{align}
To recast the holographic dictionary at finite cutoff in the standard (Dirichlet) quantization, we perform a Legendre transform with respect to the source $J$, leading to

\begin{align}\label{stdZ}
    \int DJ\ e^{-i\int J \phi_0}\langle e^{i\int JO} \rangle_{\Lambda}=\int_{r \geq r_{\epsilon}} D \phi DJ\ e^{iS_0[\phi]+iS_B-i \int J \phi_0}.
\end{align}
Here, $\phi_0$ denotes the classical source for the boundary operator $\mathcal{O}$  of dimension $\Delta$ in the dual field theory. 
This Legendre transform relates the alternate and standard quantizations by exchanging $J$ and $\phi_0$.

The left-hand side of (\ref{stdZ}) defines an effective action $I[\phi_0]$ through a Legendre transform. Given the one and two-point functions of a scalar operator $\mathcal{O}$ near equilibrium, it admits the expansion in momentum space.
\begin{align}\label{Lf}
e^{i I[\phi_0]} \approx \exp\left\{ i \int^{\Lambda} \frac{d^d k}{(2\pi)^d} \left[ \phi_0(-k) \langle \mathcal{O}(k) \rangle + \frac{1}{2} \phi_0(-k) G_s(k) \phi_0(k) \right] \right\},
\end{align}
The right-hand side of equation (\ref{stdZ}) involves integrating over the source $J$ and can be recast as
\begin{align}\label{RT}
\text{r.h.s.}_{(\text{\ref{stdZ}})} &=\int_{r \geq r_{\epsilon}} \mathcal{D}\phi \, e^{i S_0[\phi; r \geq r_{\epsilon}] + i \int_{r = r_{\epsilon}} \sqrt{-\gamma} \frac{1}{2} h \phi^2} \int \mathcal{D}J \, e^{i \int J \left( \zeta \sqrt{-\gamma} \phi(r_{\epsilon}) - \phi_0 \right)} \notag\\
&= \int_{r \geq r_{\epsilon}} \mathcal{D}\phi \, \delta\left( \phi(r_{\epsilon}) - \tilde{\phi}_0 \right) e^{i S_0[\phi; r \geq r_{\epsilon}] + i \int_{r = r_{\epsilon}} \sqrt{-\gamma} \frac{1}{2} h \phi^2},
\end{align}
where the rescaled source at the cutoff surface is defined by
\begin{align}
\tilde{\phi}_0 = \frac{1}{\zeta \sqrt{-\gamma}} \phi_0.
\end{align}
In the semiclassical approximation, the path integral is dominated by the classical solution $\phi_c$ subject to the Dirichlet boundary condition $\phi_c(r_{\epsilon}) = \tilde{\phi}_0$, yielding
\begin{align}
\mathrm{r.h.s.}_{(\ref{stdZ})} = e^{i S_c[\phi_c; r = r_{\epsilon}]} \big|_{\phi_c(r_{\epsilon}) = \tilde{\phi}_0}.
\end{align}
The on-shell action expressed in momentum space takes the form
\begin{align}\label{Rt}
S_c = \int \frac{d^{d+1}k}{(2\pi)^{d+1}} \left. \left( -\frac{1}{2} \phi_c \Pi_c + \frac{1}{2} \sqrt{-\gamma} \, \phi_c h \phi_c \right) \right|_{r = r_{\epsilon}},
\end{align}
where $\Pi_c$ denotes the radial canonical momentum conjugate to $\phi_c$.
By comparing the generating function \eqref{Lf} and the on-shell action \eqref{Rt}, we can read
off the correlation function in standard quantization
\begin{align}
G_{s} =\frac{1}{\gamma \zeta^2} \left( \frac{\Pi_c}{\phi_c} - \sqrt{-\gamma} h \right) \bigg|_{r = r_{\epsilon}}.
\end{align}
This expression is derived under the assumption that $\langle \mathcal{O} \rangle = 0$, which corresponds to a thermal equilibrium state without spontaneous symmetry breaking. 

\section{ Kullback–Leibler divergence}\label{KL}
We will show that our loss function (\ref{loss}) comes from the KL divergence of the probability distribution defined by the partition function in the field theory. The KL divergence (or relative entropy) \cite{Kullback1951} is a measure used to quantify the difference between two probability distributions. It indicates the amount of information lost when approximating one distribution with another. For two probability distributions $P_1$
 and $P_2$, the integral form of the KL divergence $\Delta_{KL}(P_1||P_2)$ is defined as
 \begin{equation}
     \Delta_{KL}(P_1||P_2)=\int_x P_1(x)\log \frac{P_1(x)}{P_2(x)} dx.
 \end{equation}

Consider the probability distribution specified by the partition function (\ref{pfft}) in Euclidean space with $\left\langle
O\right\rangle=0$:
\begin{align}
    P_i(J)=C_i \exp\left[ -\frac{1}{2}\int\frac{d\omega}{2\pi}J(\omega)G_i(\omega)J(-\omega)\right],\ \ \; \int D J\  P_i(J)=1,\ \ \;i=1,2,
\end{align}
where $G_i>0$ and the normalization factor is
\begin{equation} C_i=\det\left(\sqrt{\frac{G_i}{2\pi}}\right)=\prod_{\omega}\sqrt{\frac{G_i}{2\pi}},\ \ \;i=1,2.
\end{equation}
\( G_1 \) is interpreted as the data originating from the two-point correlation function in boundary theories. \( G_2 \) is interpreted as the holographic output generated by Neural ODEs.
In this case, the KL divergence can be rewritten as 
\begin{align}
     \Delta_{KL}&=\int_J DJ\ P_1(J)\log \frac{P_1(J)}{P_2(J)}\notag\\
    &=\int_J DJ\ C_1\exp\left[-\frac{1}{2}\int\frac{d \omega }{2\pi}JG_1J\right]\left(\log\frac{C_1}{C_2}+\frac{1}{2}\int \frac{d\omega}{2\pi}J^2(G_2-G_1)\right)\notag\\
    &=\int \frac{d\omega}{2\pi}\frac{1}{2}\left(\frac{G_2}{G_1}-1-\log\frac{G_2}{G_1}\right), \label{KL2}
 \end{align}
which motivates our choice of the loss function (\ref{loss}).

\section{Algorithm details and training report}\label{AD}
We will provide the details of the algorithm, including the setting in the architecture, hyperparameter tuning, and the training scheme, etc. We will also report the training results.

(1) Neural network

The metric function is represented by a fully connected feed-forward network. It consists of three dense layers with the following structure: \( (1, 5) \to (5, 5, \tanh) \to (5, 1) \).

(2) ODE solver

For the integration of differential equations, we employ an adaptive and explicit Runge-Kutta method, specifically the Tsitouras 5/4 solver.

(3) IR and UV cutoffs

We set the IR cutoff \( u_{IR} =0.999999\). For CFT, Lifshitz and hyperscaling-violated examples, the UV cutoffs are set as \(\epsilon=0.0001\). For the SYK model, the UV cutoff is taken as a hyperparameter.

(4) Initial values

The neural network is initialized by sampling from a normal distribution \( N(0, \sigma) \), where the standard deviation is 1 for training and 0.001 for hyperparameter tuning.

(5) Hyperparameter tuning

We use two optimizers RMSProp and Adam in sequence. Their learning rates, epoch numbers, and batch sizes are set as (0.001,0.0001), (10,10), and (1,1). Note that a total of 20 epochs is insufficient to minimize the loss, but it is adequate for obtaining the desired hyperparameters for \( CFT_1 \), the Lifshitz gravity, and the theory with hyperscaling violation. For the SYK model, this hyperparameter tuning can quickly identify the \( AdS_2 \) black hole related to the critical curve. However, one should be careful that with sufficient training, hyperparameters below the critical curve can achieve losses comparable to those along the curve.

\begin{table}[ht]
\centering
\begin{tabular}{lccc}
\toprule

 & Loss & MRE &  \\
\midrule
\textbf{CFT} & &   \\
$2\pi \bar{T}=0.6$ & $3 \times 10^{-10}$ & 0.002  \\
$2\pi \bar{T}=0.8$ & $2 \times 10^{-10}$ & 0.002  \\
$2\pi \bar{T}=1.0$ & $3 \times 10^{-10}$ & 0.002 \\
$2\pi \bar{T}=1.2$ & $2\times 10^{-10}$ & 0.002  \\
$2\pi \bar{T}=1.4$ & $2 \times 10^{-10}$ & 0.002  \\
\midrule
\textbf{Lifshitz gravity} & $3 \times 10^{-10}$ & 0.001  \\
\textbf{Hyperscaling violation} & $6\times 10^{-11}$ & 0.0001  \\
\midrule
\textbf{SYK ($v=0.99999$)} & &  \\
$2\pi \bar{T}=0.6$ & $3 \times 10^{-10}$ & 0.0008  \\
$2\pi \bar{T}=1$ & $2\times 10^{-10}$ & 0.0004  \\
$2\pi \bar{T}=1.4$ & $1 \times 10^{-10}$ & 0.0008 \\
\midrule
\textbf{SYK ($v=0.98$)} & &   \\
$(2\pi \bar{T},\epsilon)=$(0.6, 0.23) & $3 \times 10^{-10}$ & 0.009  \\
$(2\pi \bar{T},\epsilon)=$(1.0, 0.15) & $3\times 10^{-10}$ & 0.006 \\
$(2\pi \bar{T},\epsilon)=$(1.4, 0.11) & $3 \times 10^{-10}$ & 0.007  \\
\bottomrule
\end{tabular}
\caption{Minimum loss and MRE of the learned metrics in various numerical experiments.}
\label{LM}
\end{table}

(6) Training process

The training process is organized into four stages. The optimizers (RMSProp, Adam, Adam, BFGS) are applied sequentially across four stages. The learning rates for the first three stages are set to \( (0.001, 0.0001, 0.0001) \), with epoch numbers of \( (10, 10, 50) \) and a batch size of \( 1 \). In the fourth stage, the initial step size is \( 0.01 \), and the maximum iterations are \( 300 \).

In the first stage, we train \( N_1 = 50 \) times, each with randomly initialized parameters. For stages \( i > 1 \), we initialize parameters from the top \( N_i = 0.8 \times N_{i-1} \) results with the lowest loss from the previous stage. We conduct parallel training across multiple computer cores, stopping the \( i \)-th stage once more than \( 0.8 \times N_i \) tasks are completed. This allows Neural ODEs to effectively avoid getting stuck in suboptimal solutions.

After training, we select the result with the minimum loss. Using the optimal parameters, we plot the learned metric in the main text. Table \ref{LM} presents the minimum loss and the mean relative error (MRE) of the learned metric. Additionally, we gather the top 5 results with the lowest loss to construct the \( 1\sigma \) confidence intervals shown in Fig.~\ref{RTSYK98below}.

\bibliographystyle{JHEP}

\end{document}